\documentclass[aps,prl,twocolumn,showpacs,superscriptaddress]{revtex4}

\usepackage[english]{babel}
\usepackage{color}
\usepackage{amsmath, amsfonts, amssymb}
\usepackage{graphicx}

\usepackage{color}
\definecolor{Blue}{rgb}{0.00, 0.00, 1.00}
\definecolor{Red}{rgb}{1.00, 0.00, 0.00}

 \newcommand{\lch}{l_{\chi}}

 \newcommand{\Um}{\overline{U}}

 \newcommand{\Vch}{V_{\chi}}
 \newcommand{\Vf}{V_f}
 \newcommand{\Fcu}{F_{c^+}}
 \newcommand{\Fcd}{F_{c^-}}
 
 \newcommand{\Eq}[1]{Eq.~(\ref{#1})}
\newcommand{\Flau}{\hat{F}}
\newcommand{\Flcp}{\hat{F}_{c^+}}
\newcommand{\Flcm}{\hat{F}_{c^-}}
 \newcommand{\F}{F}

\begin{document}

\title{
Experimental evidence for three universality classes for reaction
   fronts in disordered flows
   }
\author{S\'everine Atis}
\affiliation{FAST, CNRS, UPSud, UPMC, UMR 7608, Batiment 502, Campus Universitaire, 91405 Orsay,
France.}
\author{Awadhesh Kumar Dubey}
\affiliation{FAST, CNRS, UPSud, UPMC, UMR 7608, Batiment 502, Campus Universitaire, 91405 Orsay,
France.}
\author{Dominique Salin}
\affiliation{FAST, CNRS, UPSud, UPMC, UMR 7608, Batiment 502, Campus Universitaire, 91405 Orsay, France.}
\author{Laurent Talon}
\affiliation{FAST, CNRS, UPSud, UPMC, UMR 7608, Batiment 502, Campus Universitaire, 91405 Orsay,
France.}
\author{Pierre Le Doussal}
\affiliation{CNRS-Laboratoire de Physique Th\'eorique de l'Ecole Normale Sup\'erieure, 24 rue Lhomond, 75005 Paris, France.}
\author{Kay J\"org Wiese}
\affiliation{CNRS-Laboratoire de Physique Th\'eorique de l'Ecole Normale Sup\'erieure, 24 rue Lhomond, 75005 Paris, France.}

\begin{abstract}
Self-sustained reaction fronts in a disordered medium subject to an external flow display self-affine roughening, pinning and depinning transitions. We measure spatial and temporal fluctuations of the front in $1+1$ dimensions, controlled by a single parameter, the mean flow velocity. Three distinct universality classes are observed,  consistent with the Kardar-Parisi-Zhang (KPZ) class for fast advancing or receding fronts, the quenched KPZ class (positive-qKPZ) when the mean flow  approximately cancels the reaction rate, and the negative-qKPZ class for slowly receding fronts. Both quenched KPZ classes exhibit distinct depinning transitions, in agreement with the theory.

\end{abstract}

\maketitle

Growing interfaces are ubiquitous in nature, appearing in situations as different as bacterial colonies \cite{huergo10}, solidification \cite{langer80}, atomic layer deposition \cite{messier85, wakita97}, liquid interfaces in porous media \cite{wilkinson83,horvath91,santucci11} or crack propagation in heterogeneous materials \cite{bouchaud97,maloy01}. The formation of scale-free structures in these systems raises the important question of universality in out-of-equilibrium phenomena. 

In this letter, we consider the propagation of a reaction front inside a porous medium. Resulting from the balance between the molecular diffusion $D_m$ and the reaction rate $\tilde \alpha$, autocatalytic reactions can develop a traveling front.  In the absence of an externally imposed flow, the reaction front develops into a flat horizontal front, propagating with a constant velocity $V_{\chi}=\sqrt{D_m { \tilde \alpha} /2}$, and a stationary concentration profile of width $l_{\chi}=D_m /V_{\chi}$. When coupled with the heterogeneous flow field of the porous medium, the fronts become rough, and modify their behavior accordingly with the flow strength and mean orientation relative to the chemical reaction direction. They propagate either downstream or upstream, or can remain frozen over a range of counter-flow rates, delimited by two distinct depinning transitions \cite{atis13,saha13}.
Until now, however, their universality classes have not been identified.

\begin{figure}[tb]
\includegraphics[width=8.6cm]{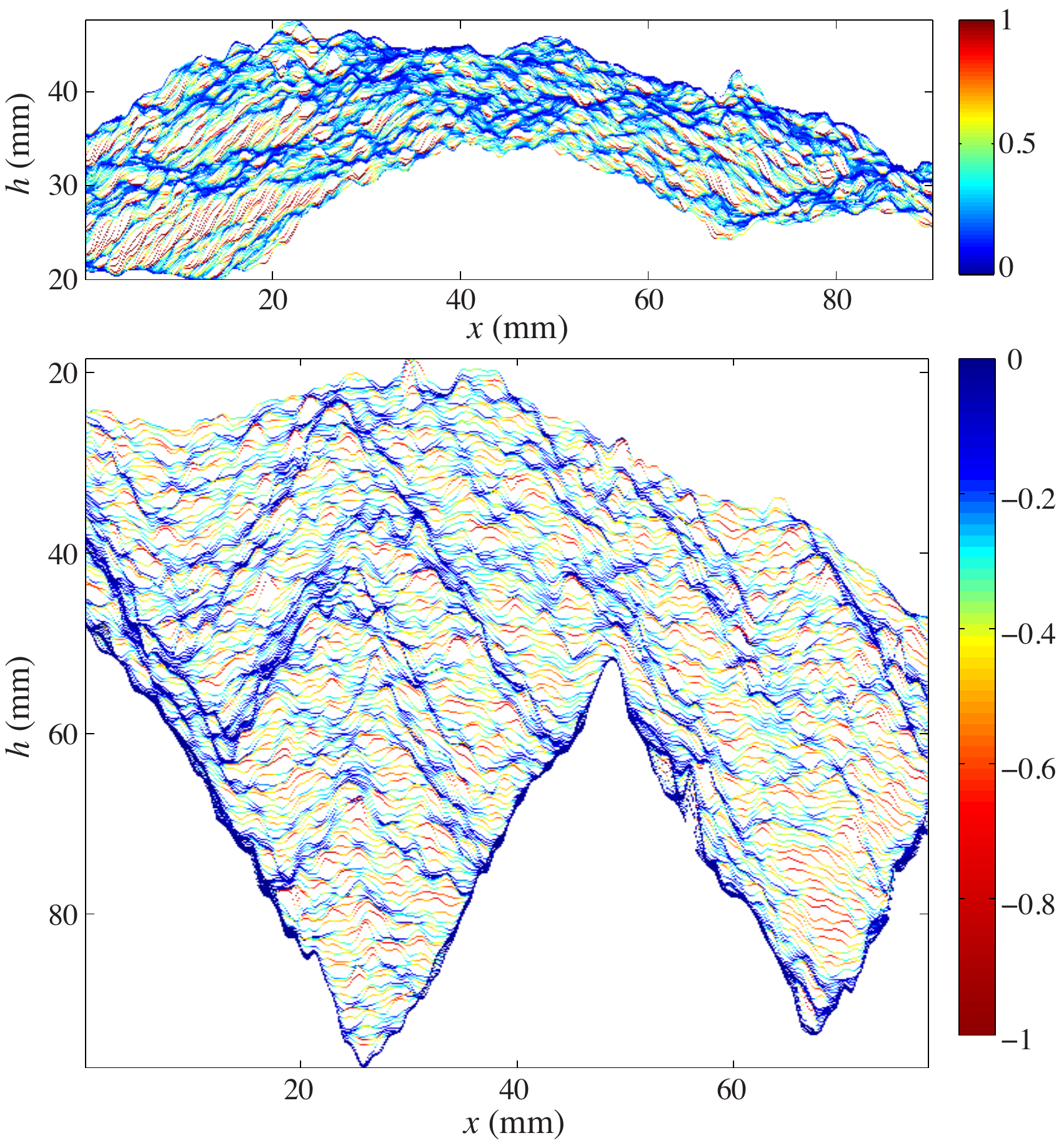}
\caption{Successive experimental fronts at constant time intervals. Color represents local front velocity, in units of imposed flow velocity $\bar U$. Top:
upward propagating front near $F_c^+$ ($F=0.56$). Bottom: backward propagating front near
$F_c^-$ ($F= - 1.24$).}
\label{field}
\end{figure}

Using both experimental and numerical approaches, we investigate their spatial and temporal scaling
over the whole range of the externally imposed flow.
In the vicinity of both depinning points, these reaction fronts display transient static configurations with distinct morphologies depending on the front propagation direction, displayed  on Fig.~\ref{field}.
We show that this is a well-controlled system which encompasses several universality classes.

Two important classes predicted by the theory, and discussed below, are:
\\
(i) non-linear stochastic growth governed by the (thermal) Kardar-Parisi-Zhang (KPZ) equation (\ref{KPZstoch}),
\\
(ii) growth where both the non-linearity and quenched disorder are present, described by the quenched KPZ (qKPZ) equation (\ref{qKPZ}). It divides into two subclasses, positive qKPZ and negative qKPZ, depending on the sign of the non-linearity $\lambda$. 
It has been difficult to find unambiguous experimental realizations,  due to long-range effects, quenched disorder, and a mixing of (i) and (ii)  \cite{maunuksela97,takeuchi10,huergo10,yunker13}. Recently,  experiments on turbulent liquid crystals \cite{takeuchi10, takeuchi14} made a precise connection with the theory of the KPZ class experiencing a revolution of its own \cite{corwin}. The (positive) qKPZ class \cite{buldyrev92,amaral95} and both KPZ and qKPZ classes in evaporating colloidal suspensions \cite{yunker13} were observed. Remarkably, in the present system, by tuning a {\em single} parameter $F$, one can observe {\em all three classes}. 

\begin{figure}[tbt]
\includegraphics[width=8.5cm]{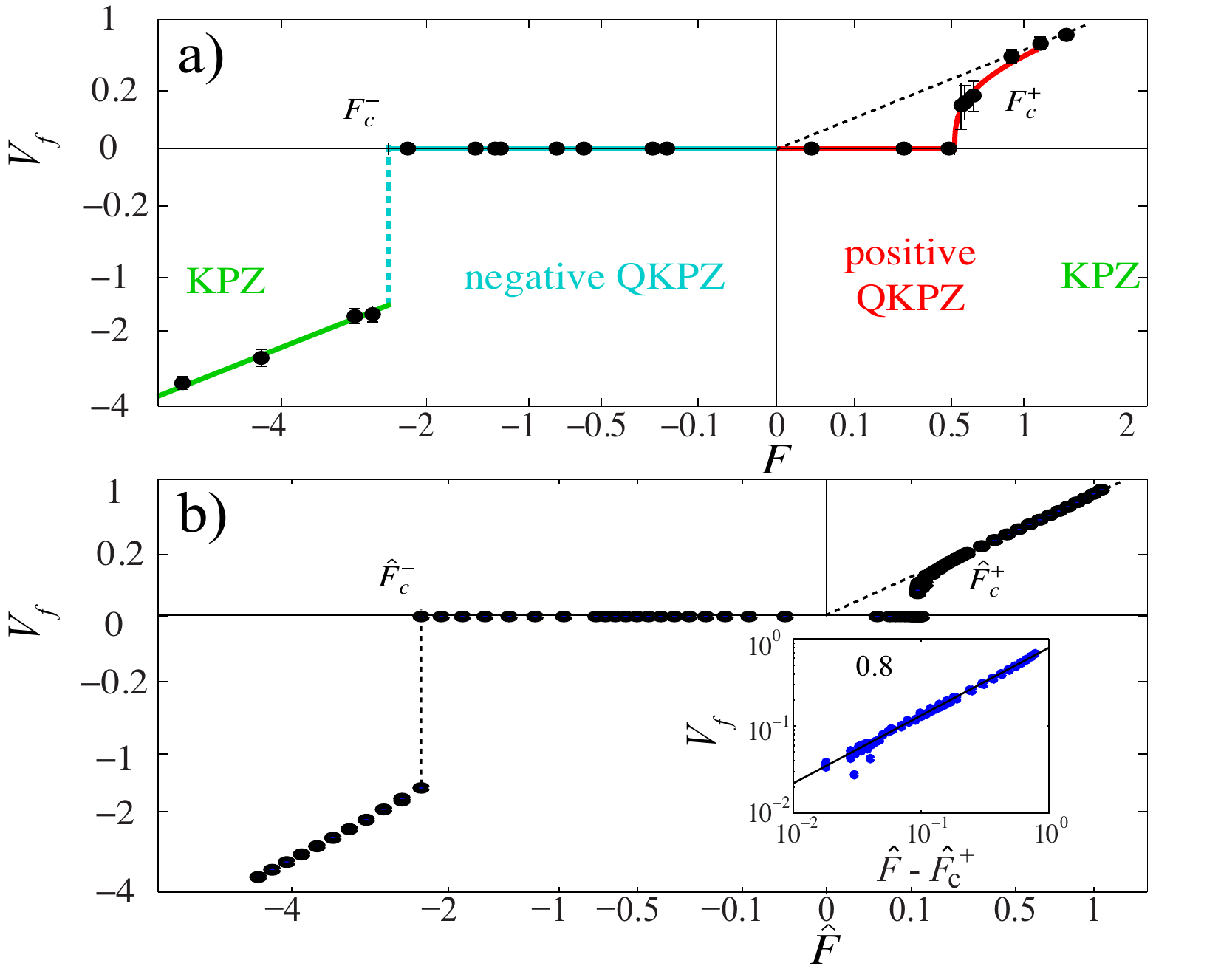}
\caption{Front velocity $V_f$ versus the applied force $F$ (resp. $\hat F$), in adverse flow 
a) experiments (black dots with error bars), b) numerics. Dashed lines are linear extrapolation of the advancing branch.
To put all data on one plot, axes are rescaled according to $F\to F/|F|^{1/2}$, $V_f\to V_f/|V_f|^{1/2}$.
Insert: front velocity versus $\Flau - \Flcp$. The continuous line corresponds to
$v(\hat F) \propto ( \hat F - \hat F_{c^{+}} )^{ 0.8 \pm 0.05}$.}
\label{VF} 
\end{figure}

The KPZ equation \cite{kardar86} was proposed as a generic model for an interface growing along its local normal,\begin{equation}\label{KPZstoch}
 \frac {\partial h(x,t)}{ \partial t}  = \nu \nabla^2 h(x,t) + \frac{\lambda}{2}\left[ \nabla h(x,t)\right]^2 +\eta(x,t) + f\;
.\end{equation}
Its height $h(x,t)$ is along the vertical axis, $\nu$ an effective stiffness due to diffusion, $\lambda$ the non-linearity, and $\eta(x,t)$ a Gaussian white noise with $\overline{\eta(x,t)}=0$ and $\overline{\eta(x,t)\eta(x^\prime,t^\prime)} = 2D\delta(x-x^\prime)\delta(t-t^\prime)$. $f$ is an applied force, and up to a shift, proportional to the experimental applied force $F$ as shown below. The surface can be characterized by two scaling exponents, the roughness $\alpha$, and the  growth exponent $\beta$, defined via  $\overline{[h(x,t)-h(x',t)]^2}\sim |x-x'|^{2\alpha}$ and $\overline{[h(x,t)-h(x,t')]^2}\sim|t-t'|^{2\beta}$.
In $d=1+1$ dimensions, $\alpha_{\mathrm{KPZ}}=1/2$ and $\beta_{\mathrm{KPZ}}=1/3$ \cite{kardar86}.

\begin{figure}
\includegraphics[width=8.6cm]{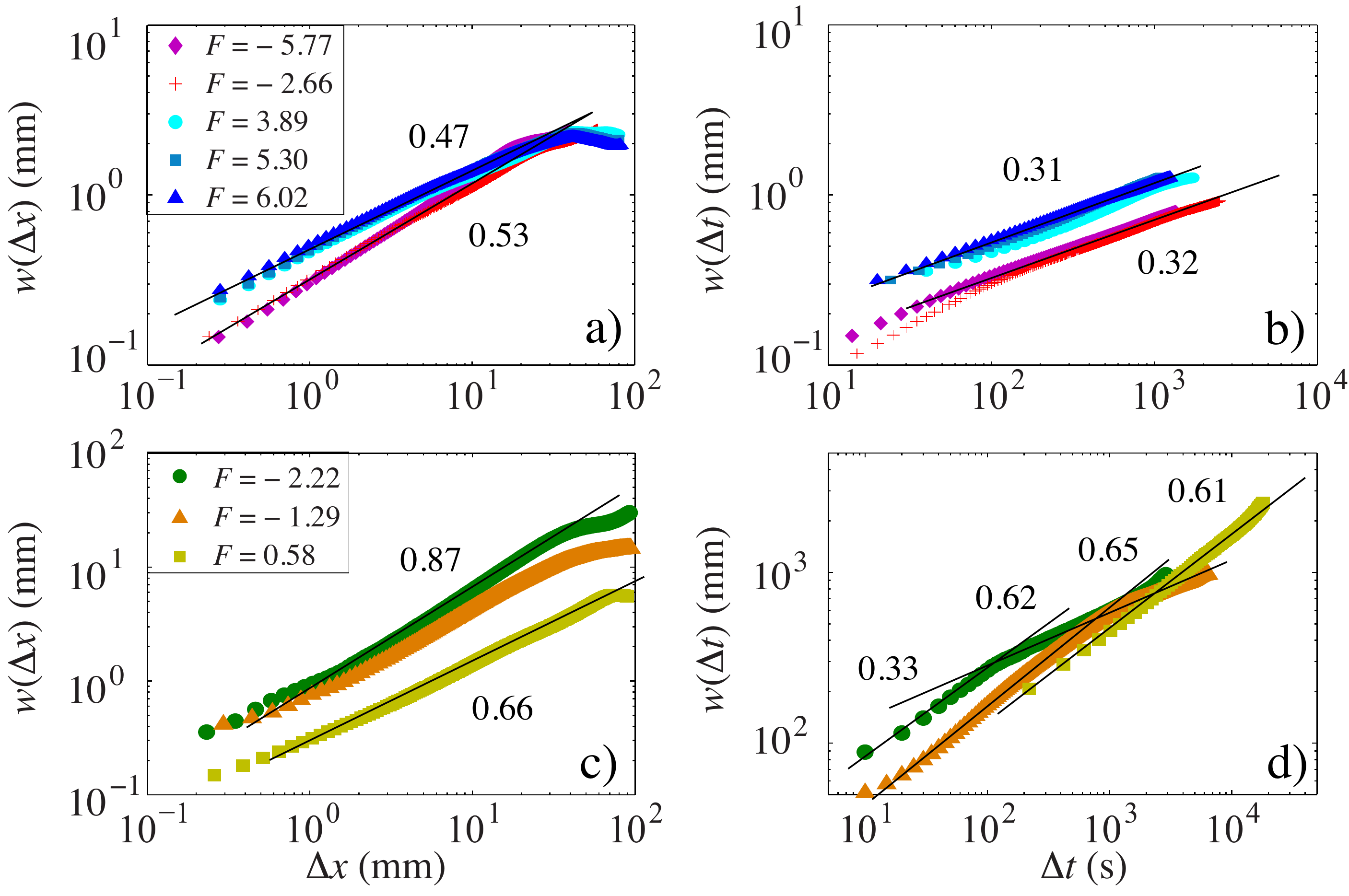}
\caption{Height fluctuations of the front. Left column: roughness $w(\Delta x)$ and right column: temporal fluctuations $w(\Delta t)$. (a)~and (b)~for $|F|>2.5$; (c)~and (d) for $|F|<2.5$.}
\label{W}
\end{figure}
In a heterogeneous medium, the ``noise" acquires a static {\em quenched} component, described by the qKPZ equation \cite{kessler91}:
\begin{equation}\label{qKPZ}
\frac {\partial h(x,t)}{ \partial t}  = \nu \nabla^2 h(x,t) + \frac{\lambda}{2}\left[ \nabla h(x,t)\right]^2 +\bar \eta\Big(x,h(x,t)\Big) + f\;.
\end{equation}
The case $\lambda=0$ models a number of systems, and is a distinct universality class (quenched Edwards-Wilkinson) \cite{QEW} but does not seem to be relevant here (it predicts $\beta \approx 0.87$ and $\alpha>1$.)
In the KPZ Eq.\ (\ref{KPZstoch}) one can eliminate the driving by setting $h(x,t) = f t + \tilde h(x,t)$.
Changing then $\tilde h(x,t) \to -\tilde h(x,t)$ reverses the sign of the non-linear term $\lambda$, which is thus unimportant. By contrast, the qKPZ Eq.~(\ref{qKPZ}) does not allow for this change, since the term $\bar \eta(x,h(x,t))$ is not invariant. The driving force $f$ is thus a new parameter of the problem, and its sign (relative to the sign of $\lambda$) matters. If the disorder is statistically invariant by parity, i.e.\ if $\bar \eta(x,-h)$ has the same properties as $\bar \eta(x,h)$, \Eq{qKPZ} is invariant under $f\to -f$,  $h(x,t)\to -h(x,t)$, and $\lambda \to - \lambda$, leading to two cases: Positive qKPZ when $\lambda$ and $f$ have the same sign, and negative qKPZ when they have opposite signs.

In the moving phase, qKPZ of either sign crosses over to KPZ at large scales. This is seen e.g.\ in the limit of large mean interface velocity $v= \overline{\partial_t h(x,t)}$: Consider \Eq{qKPZ} with white noise $\overline{ \bar\eta(x,h) \bar \eta(x',h') }= 2 \bar D \delta(x-x') \delta (h-h')$ and perform the change $h(x,t) \to v t + \tilde h(x,t)$. The disorder then becomes
\begin{equation} \label{eq3}
\bar \eta \big(x, v t + \tilde h(x,t) \big) \approx \bar \eta (x,v t) \;,
\end{equation}
i.e.\ the same noise as in the KPZ Eq.\ (\ref{KPZstoch}), identifying $D=\bar D/v$. As $v$ is decreased, the crossover from qKPZ at short scales to KPZ occurs at larger and larger scales.

The positive qKPZ equation exhibits a depinning transition \cite{kessler91,parisi92}, well characterized in $d=1+1$. The mean interface velocity vanishes for $f< f_{\rm c}^+$; for $f> f_{\rm c}^+$ it moves with velocity $v \sim (f-f_{\rm c}^+)^\theta$. At $f_{\rm c}$ the pinned interface outlines 
a transversal path on a directed percolation (DP) cluster \cite{buldyrev92,tang92,amaral95} with roughness $\alpha_{\mathrm{DP}} \simeq 0.63$, and a growth exponent $\beta_{\mathrm{DP}} \simeq 0.63$. As discussed above the quenched nature of the noise is relevant
only when $f$ is close to $f_{\rm c}$.

The predictions for the negative qKPZ class are different \cite{coreens}. In the pinned phase, $\lambda>0$ and negative $f>f_{\rm c}^-$, the interface forms sawtooth configurations (see Fig.~\ref{field}, bottom) where alternating non-zero local average slopes $|\nabla h|$ help the system to remain pinned.  As $f$ decreases, the sawtooth slopes increase until there are discontinuous jumps at $f_{\rm c}^-$ of both the average slope (back to zero) and the velocity $v$, well evident in our experiment, see Fig.~\ref{VF} a). Inside the pinned phase, the transitory dynamics is similar to positive qKPZ \cite{coreens2}, and the depinning was analyzed via a mapping to the first layer PNG model \cite{szabo01}, a close cousin of KPZ.

Our experiment is made with the Iodate Arsenous Acid (IAA) reaction, autocatalytic in iodide, with the concentration values $[\mathrm{IO}_3^-] = 7.5$ mM and $[\mathrm{H}_3\mathrm{AsO}_3]_0 = 25$ mM, such that the arsenous acid is in excess \cite{hanna82,leconte03}. 
The resulting front has a velocity $V_{\chi}\simeq 11.2\; \mathrm{\mu m/s}$, and a width $l_{\chi}\simeq 200 \;\mathrm{\mu m}$.
The front position is visualized with polyvinyl alcohol colored by transient iodine production  \cite{boumalham10}.
The disordered flow is generated with a $50\%$ mixture of $1.5$ and $2 $ mm diameters packed glass beads inside a transparent ($300 \times 100 \times 4 \, \mathrm{mm^{3}}$) rectangular cell. A range of injectors at the top of the cell can either suck out or inject unreacted fluid parallel to the vertical. The bottom of the cell is dipped into a container with reacted solution to start the reaction, leading to a flat horizontal initial front. The front propagates upwards in the absence of the flow, which is then switched on once the desired vertical position is reached. We extract the location $h(x,t)$ of the front at a given position $x$ and time $t$ from $700\times1000$ pixel digitized images with a 12-bit Roper Coolsnap HQ video camera. To enhance the statistics, we performed 3 to 6 different realizations for each of the 30 values of the applied flow velocity.

We define the control parameter as $F = (\Um + V_\chi)/V_\chi + f_0$, whith $\Um$ the mean flow velocity and $f_0\simeq0.38$ an \emph{ad-hoc} constant \cite{SM}, such that the front advances when $F>0$ or recedes when $F<0$. Fig.~\ref{VF} a) displays the normalized front velocity, $v=V_f/V_{\chi}$ as function of $F$.
\begin{figure}
\includegraphics[width= 8.7cm]{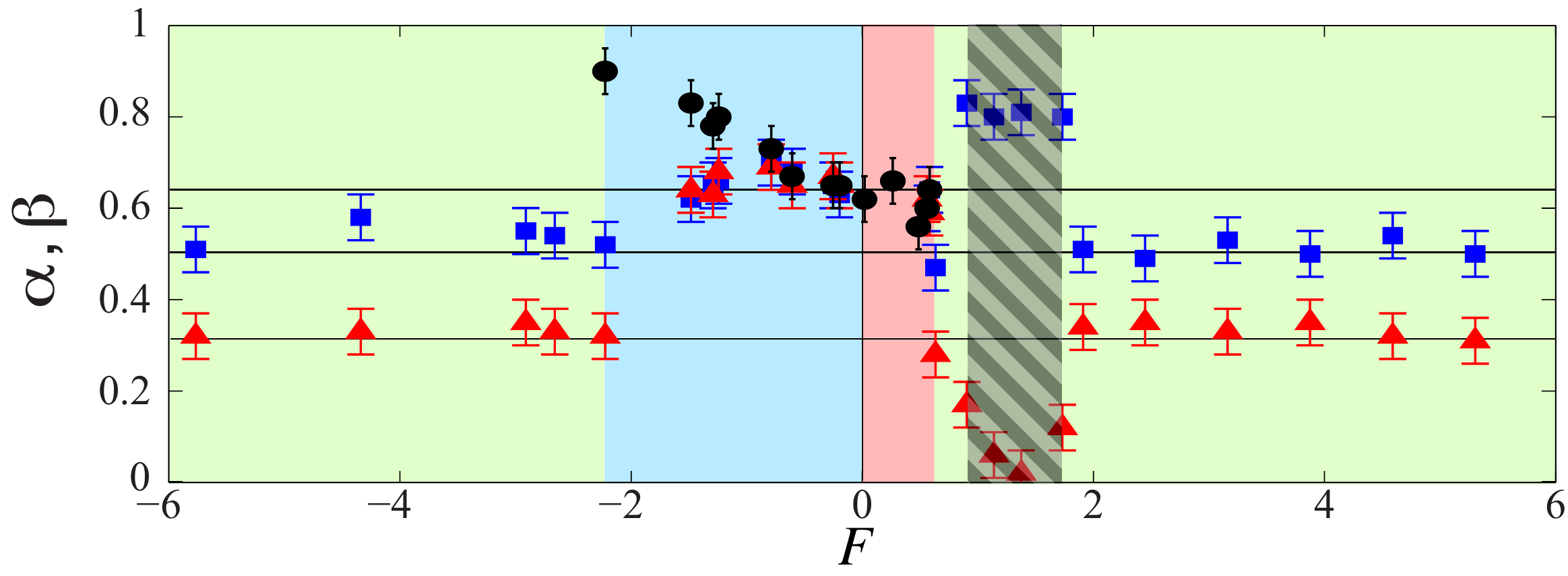}
\caption{Experimental exponents versus $F$. Roughness $\alpha$ (blue squares), and  growth exponent $\beta$ (red triangles) for moving fronts. Roughness of the pinned fronts (black circles). Hatched region corresponds to weak noise  when $\Um\rightarrow 0$.}
\label{H_beta}
\end{figure}
In the absence of flow, when $F=1+f_0$, the reaction front propagates through the beads with a smooth shape at a constant velocity $V_f=V_0 = 0.8 \Vch \pm0.5 \;\mu$m/s [supp. movie 1 in \cite{SM}]. When the flow is turned on, the front exhibits distinct self-affine scalings depending on $F$, and quantified by the front width $w(\Delta x)\sim (\Delta x)^{\alpha}$, and the standard deviation of the temporal height fluctuations $w(\Delta t)\sim (\Delta t)^{\beta}$ \cite{SM,footnote1}.

At large flow rates, when $|F| > 2.5$ on Fig.~\ref{VF}, $V_f$ is a linear function of $F$ and the front propagates downstream for both orientations of the mean flow [supp. movie 2 and 3 in \cite{SM}]. Fig.~\ref{W} a) shows the width $w(\Delta x)$ and $w(\Delta t)$ of saturated fronts,  i.e.\ such that $\ell^*(t) \sim L$ \cite{footnote1}, determined for experiments realized at opposite mean-flow orientations. They display both a similar roughness  $\alpha = 0.47\pm0.03,$ and $\alpha = 0.53\pm0.04,$ and growth exponent $\beta= 0.32\pm0.04$ and $\beta= 0.37\pm0.05$ while $t^* \approx T$. As can be seen on Fig.~\ref{H_beta} a), at large front velocity of either orientation, the front exhibits scale-invariant fluctuations with statistical properties in agreement with the KPZ class and the theoretical discussion around Eq.~(\ref{eq3}). In addition, since in the experiment $ \bar D\sim \Vf$, the expected KPZ noise $D \simeq \bar D/\Vf$ is almost independent of $\Vf$ \cite{SM}, and the amplitude of $w(\Delta x)$ does not vary significantly with $F$ for a given flow orientation, as can be seen in Fig.~\ref{W} a).

When $F \rightarrow 0$, some regions of the front pin to the flow heterogeneities. In this configuration, the front propagates mainly upstream, from the bottom to the top of the cell, while locally the front exhibits transiently static regions, as shown in Fig.~\ref{field} (top). Note that the moving parts have a larger slope than the arrested or slowly propagating ones, leading to a lateral growth of the fronts in this regime (supp. movie 4 in \cite{SM}). When the opposite flow is amplified, the pinned portions become larger. The value $\Fcu$ for which the front eventually stops and remains static is $\Fcu = 0.56\pm0.05$.
On Fig.~\ref{W} c) and d) for $F=0.58$, the values $\alpha = 0.66\pm0.04$ and $\beta = 0.61\pm0.05$, are consistent with the theoretically predicted exponents of positive qKPZ, $\alpha=\beta=0.63$ ({see \cite{SM} for additional measurements}), suggesting that the front undergoes a depinning transition when $F \rightarrow \Fcu$.

Finally, when $F$ decreases below $\Fcu$, the transient front propagation becomes very short. For $F\approx 0$, the front is static almost instantaneously after the flow is turned on (supp. movie 5 in \cite{SM}). When $F$ becomes negative, the front propagates in the direction opposite to the chemical reaction. For sufficiently small $F$, $-2.22 \lesssim F\lesssim 0$, it quickly becomes static after a transient propagation and displays a particular sawtooth pattern \cite{atis13}. One notes in Fig.~\ref{field} b) that the front is slowed down or arrested when it reaches a certain slope, resulting in facet formation (supp. movie 6 in \cite{SM}).
Another depinning transition occurs at $\Fcd \approx -2.22\pm0.05$, below which triangular states become unstable and the front goes back to a  phase moving from the top to the bottom of the cell.
As indicated in Fig.~\ref{H_beta}, these triangular shapes lead to a different roughness exponent of the transiently moving parts compared to the final static front as $F\rightarrow\Fcd$ \cite{SM}. The roughness and growth exponents of propagating regions have $0.62 \lesssim\alpha\lesssim 0.7$ and $0.63 \lesssim\beta\lesssim 0.69$, consistent with the observations in \cite{coreens2}.
However, the final static front roughness is larger: $0.73 \lesssim\alpha\lesssim 0.9$, increasing as the sawtooth slope rises when $F \rightarrow \Fcd$.
Interestingly, a crossover from $\beta \approx 0.65$ to $\beta \approx  0.33$ for larger scales is visible in Fig.~\ref{W} d), underlining the second depinning transition at $\Fcd$. Close to $\Fcd$, the front pins to point-like regions, while close to $\Fcu$ the pinning regions extend horizontally. This shows  that receding fronts are consistent with negative qKPZ,  known for similar pinning processes and interface morphologies \cite{coreens,szabo01,moon2013}. This model also predicts a first-order depinning transition, observed here as a jump in the $\Vf(F)$ curve.

\begin{figure}
\includegraphics[width= 7.8cm,trim=50 0 50 0]{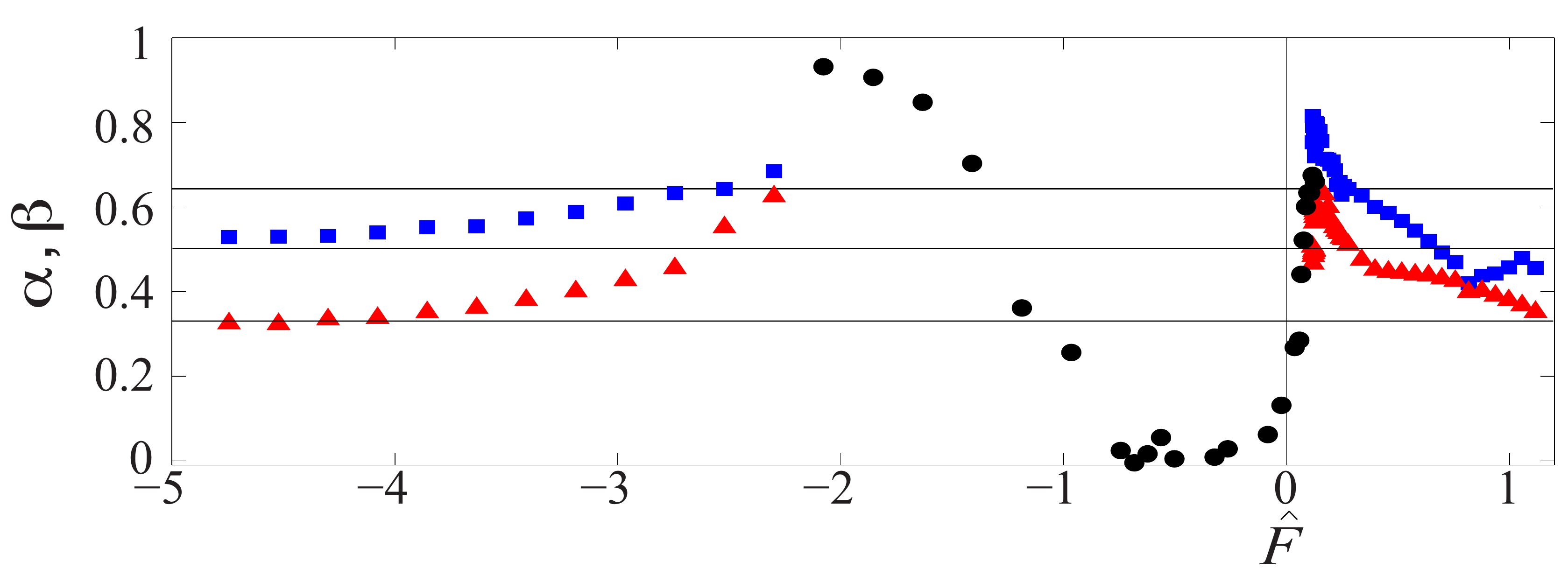}
\caption{Roughness and growth exponents in the simulations. Same symbols as in Fig. \ref{H_beta}}
\label{H_betasim}
\end{figure}
To better understand the behavior close to the transitions, lattice Boltzmann simulations were  performed in a $2D$ disordered porous medium ($2048\times2048$ grid size), solving the convection-diffusion reaction and the Darcy-Brinkman equations \cite{talon03,saha13,SM}. Fig.~\ref{VF} b) displays the numerical front velocity versus $\Flau = (\Um+ V_\chi)/V_\chi + \hat f_0$, where hatted quantities denote parameters in the simulations with $\hat f_0=0.256$. In good agreement with the experiments, two transitions occur at $\Flcm = -2.2 \pm 0.2 $ and $ \Flcp = 0.095 \pm 0.015$ and the velocity $\Vf(\Flau)$ is almost linear in $\hat F$ below the first transition and beyond the second one. While the transition at $\Flcm$ is very abrupt, the second one at $ \Flcp$ is more continuous. Moreover, the critical behavior can be fitted (Fig.~\ref{VF}.b, insert) with a power-law $\Vf(\hat F) \sim ( \hat F - \Flcp )^{0.8 \pm 0.05} $, with an exponent slightly larger than the theoretical value $0.66$ \cite{amaral95}.
This is consistent with a first-order transition at $\Flcm$ and a second-order one at $\Flcp$. Shown on Fig. \ref{H_betasim}, the scaling exponents of the numerics are also in good agreement with the theory of the thermal KPZ class for large $|\Flau|$. Near the depinning transitions at $\Flcp$ and $\Flcm$, the roughness exponents $\alpha=0.65 \pm 0.05$ and $\alpha=0.9 \pm 0.05$ are consistent with experiments and the positive and negative q-KPZ predictions. The remaining differences in the q-KPZ pinned phase are due to different initial conditions: In the experiment, the fronts propagate without flow (with their own roughness exponent at $\Um=0$) and then the flow is switched on, whereas in the simulation the initial front is flat (see \cite{SM} for details). 

The good agreement between experiments, numerical simulations, and  theory for the different KPZ universality classes can be understood through the {\em eikonal approximation} \cite{edwards02,edwards06}. For a thin front, the local front velocity follows the eikonal equation:
$\vec \Vf \cdot \vec n =  \Vch + D_m \kappa + \vec U(\vec r) \cdot \vec n$, where $\vec n$ is the normal of the front, $\kappa$ the curvature and $\vec U(\vec r)$ the local flow velocity. Indeed, this equation is similar to the ``flux-line model'' of Kardar \cite{kardar98}
where the chemical velocity plays the role of the Lorentz force, and the
disordered flow that of the random force. After projection and neglecting higher-order terms \cite{SM}, the eikonal approximation  yields
\begin{equation}
 \frac{\partial h}{\partial t} \simeq  V_\chi\sqrt{1  + (\nabla h)^2} + \frac{D_m \nabla^2 h} {1  + (\nabla h)^2} + \Um + \delta U_y(\vec r).
\end{equation}
Assuming small gradients, and normalizing by $V_\chi$, leads to Eq.~\eqref{qKPZ} where $\bar \eta \equiv \delta U_y$ and with the parameters $\nu  =  l_\chi = D_m / \Vch $, $\lambda  =  1$, and $f  =  ({\Um + \Vch })/{\Vch}$, whose small renormalization due to the neglected terms can be estimated \cite{PLD-KW-QKPZ}.
The difference $F - f = f_0$ is related to the space average of the KPZ term $f_0 \sim \frac{\lambda}{2} \left< (\nabla h)^2 \right>_L$.
Note that $\lambda=1$ is independent of the front propagation direction, and fixed by the initial condition of the experiment. Negative qKPZ describes then the backward moving fronts, i.e. $\partial h / \partial t <0$, since performing $h \rightarrow -h$ is equivalent to measuring the front position along the $-\hat{y}$ axis. Finally, near the transition at $\Fcd$, the slope of the sawtooths may be large. Although it correctly predicts the first-order transition, small-gradient qKPZ may not be quantitatively accurate. A  more precise scenario was proposed in \cite{gueudre14}, based on the PNG model \cite{szabo01} and extreme-value statistics.

In conclusion, chemical-wave propagation coupled with the disordered flow in a porous medium, develop self-affine structures, with scaling exponents consistent with either KPZ or qKPZ classes. Remarkably, by tuning a single parameter, this system passes through three universality classes, providing a rich experimental setting to study growth phenomena. Slowly backward propagating fronts constitute beautiful experimental evidence of a chemical interface described by negative qKPZ. Part of this phenomenology was recently observed in magnetic domain walls \cite{moon2013}: it would be interesting to reach the thermal KPZ class there by increasing the driving. This work opens the door for further investigations on frozen pattern formation in out-of-equilibrium systems \cite{schaller11}. 

\acknowledgments This work was supported by Project Procathet RTRA Triangle de la Physique and  PSL grant ANR-10-IDEX-0001-02-PSL. We are grateful to K. Takeuchi, T. Gueudre and A. Rosso for useful discussions.


\newpage
\thispagestyle{empty}
\mbox{}

\begin{widetext}

\textbf{Supplemental Material for ``\em Experimental evidence for three universality classes for reaction fronts in disordered flows''} 

\section{I.\;\;\;\;Experimental details}

Depending on the mean front  velocity,  between $200$ and $2000$ photos are recorded for each choice of parameters. To enhance the statistics in the transient propagation regime, for $ - 2.22 \lesssim F\lesssim 0.58 $, we performed between $3$ and $6$ different realizations for each value of $F$. A total of 30 different values of $F$ have been investigated, spanning from $-6$ to $8$.

\begin{figure}[b]
\includegraphics[width=16cm]{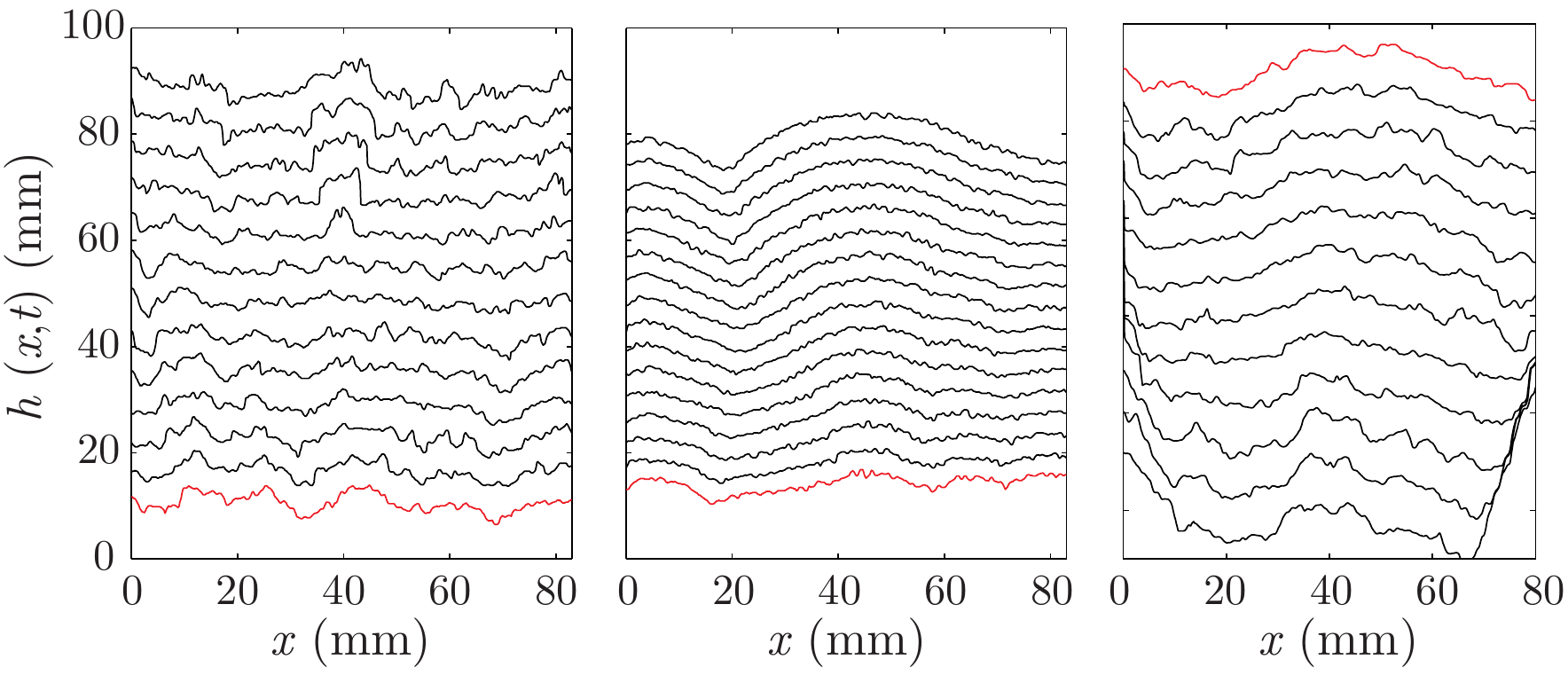}
\caption{Successive experimental fronts at constant time intervals (red: starting front). Left: $F=6.01$, upward front propagation with supportive flow. Center: $F=1.38$: relaxation pattern in the absence of flow. Right: $F=-2.66$, backward propagating front in adverse flow.}
\label{fronts}
\end{figure}
Example of the extracted fronts are given on Fig.~\ref{fronts}.
In the absence of the flow ($F=1.38$), the reaction fronts are uniformly moving through the glass beads with a constant velocity $V_f=V_0 = 0.8 \Vch \pm0.5 \;\mu$m/s and $\Vch = 11.2\;\mu\text{m/s}$, such that initially rough fronts gradually relax toward a smooth front shape [see movie 1]. For a high flow strength ($|F| \gtrsim 2.5$), the fronts become rough and present self-affine structures [see movie 2 and 3].

\section{II.\;\;\;\;Numerical methods}

The heterogeneous permeability field  $K(x,y)$ is generated according to a log-normal distribution, which is Gaussian correlated over a length $l_K$, as commonly assumed to model groundwater flow \cite{geostat}.
The flow is computed by solving the Darcy-Brinkman equation
\begin{equation}
\vec 0 = f_{b} \vec e_x - \frac{\mu}{K(x,y)} \vec U(x,y) + \mu \Delta \vec U(x,y)\;\text{,}
\end{equation}
where $\mu$ is the fluid viscosity and  $f_{b}$ is the body force driving the flow (\emph{e.g.} the buoyancy). The boundary conditions are periodic in the lateral  direction. A convection-diffusion-reaction equation is then solved for the concentration $C(x,y,t)$ of the auto-catalytic reactant (iodide), normalized by the initial concentration of iodate,
\begin{equation}
\frac{\partial C}{\partial t} + \vec U(x,y). \vec \nabla C = D_m \Delta C + \tilde \alpha C^2(1-C)\;\text{.}
\end{equation}
Here $D_m$ and $\tilde \alpha$ are  the molecular diffusion and the reaction rate. Both equations are solved using Lattice Boltzmann schemes (see \cite{talon03,ginzburg08c,saha13} for details). In the numerical study, the parameters $\tilde \alpha$ and $D_m$ have been chosen to keep constant the ratio $\lch/l_K = \frac{ \sqrt{2 D_m} } {l_K \sqrt{\tilde \alpha}} = 0.126$, and the standard deviation of the log-normal distribution has been kept constant $\sigma_{\ln K} =0.5$. As in the experiments, the control parameter $F$ is determined from the relative velocity between the reaction velocity $\Vch$ and the mean-flow velocity $\Um$. The grid size is $2048\times2048$ and the computational cost per realisation is around $10$ hours using $16$ CPUs on a standard workstation.

For each value of $F$, around four different realizations were performed, and over  800 fronts are computed for each simulation. A total of hundred different values of $F$ have been numerically generated, with thirty points close to the transition point at $\Fcu$.

\section{III.\;\;\;\;Front width determination}


Rough interface fluctuations are generally characterized by the standard deviation of their height over a length scale $\Delta x$, defining a roughness exponent $\alpha$. In the presence of ``normal scaling'' (see \cite{ramasco00,RossoKrauth2001b} and section \ref{s:sawtooth} below for exceptions) one has the choice between different methods. While  we used several ones, a common choice,  and the one employed in the data presented in the present work is to measure
\begin{equation}
w(\Delta x,t) = \left<\sqrt{\left< [h(x,t)-\left<h\right>_{\Delta x}]^2\right>_{\Delta x}}\right>_x\ ,
\end{equation}
 where $\left< ... \right>_{\Delta x}$  correspond to an average over the interval $\Delta x$ and   $\left< ... \right>_x$ is an average  over the whole interface.
In the context of kinetic roughening, an initially flat front develops power-law distributed structures as a function of time $t$, which in the absence of a  length scale in the system implies the  Family-Vicsek scaling \cite{family85}
\begin{eqnarray}
w(\Delta x,t) &\sim& t^\beta \mathcal{F}(\Delta x/t^{1/z})\;\text{,}\\
\text{with the scaling function } \mathcal{F}(u) &\sim& \left\{
\begin{array}{lcr}
u^\alpha    &\text{  if  }&u\ll 1 \;\text{,}\\
\text{const}&\text{  if  }& u\gg 1 \;\text{.}
\end{array}
\right.
\end{eqnarray}
The exponents $\alpha$ and $\beta$ are the roughness and growth exponents. The exponent $z=\alpha/\beta$ is the dynamic exponent and characterizes the lateral correlation length growth $l(t)^* \sim t^{1/z}$ for short time, before the front fluctuations reach the steady regime.
In this scenario, one can measure the growth exponent $\beta$ by determining the overall front width $W(L,t) = \sqrt{[h(x,t)-\left<h\right>_x]^2}\sim t^\beta$. This scaling requires that the initial front starts as a straight line. Although this has been done in some experiments \cite{takeuchi10, huergo10},  it is generally hard to achieve experimentally: As the reaction is propagating through the glass beads, the initial fronts usually acquire a non-trivial shape as illustrated on  Fig~\ref{fronts}.
Instead, one can use the front fluctuations in the steady state to determine the scaling exponents \cite{kardar98,ponson06}. This is achieved by determining the local width for both spatial and temporal fluctuations in the stationary regime 
\begin{eqnarray}
w(\Delta x,t) &=& \left<\sqrt{\left< [h(x,t)-\left<h\right>_{\Delta x}]^2\right>_{\Delta x}}\right>_{x}\sim \Delta x ^\alpha\;\text{,}\\
w(\Delta t,x) &=& \left<\sqrt{\left< [h(x,t)-\left<h(t)\right>_{x}]^2\right>_{\Delta t}}\right>_{t}\sim \Delta t ^\beta\;\text{.}
\end{eqnarray}
In order to improve the statistics, an additional average can be made in the stationary regime over a  time window and space window respectively to determine $w(\Delta x)$ and $w(\Delta t)$, 
\begin{eqnarray}\label{set1}
w(\Delta x) &=& \left< w(\Delta x,t) \right>_t \\
\label{set1'}
w(\Delta t) &=& \left< w(x,\Delta t) \right>_x
\end{eqnarray}
Alternatively,  the fronts can be  characterised  by the 2-point correlation functions in the steady state
\begin{eqnarray}\label{set2}
&&\left< [h(x,t)-h(x+\Delta x,t)]^2 \right>_{x,t} \sim \Delta x^{2\alpha}\\
&&\left< \left[h(x,t+\Delta t)-h(x, t) -v \Delta t \right>]^2 \right>_x \sim \Delta t^{2\beta} \\
&&  v \Delta t := \left< h(x,t+\Delta t)-h(x, t) \right>  \label{set2'}\ .
\end{eqnarray}
We found no significant differences in the estimates of the critical exponents when using either Eqs.~(\ref{set1})-(\ref{set1'}) or (\ref{set2})-(\ref{set2'}).

%

\section{IV.\;\;\;\;Separation of the scaling between the static sawtooth shapes and the propagating regions of the fronts}
\label{s:sawtooth}

The front displays static sawtooth shapes at large scales when $F$ becomes close to $\Fcd$, in the negative qKPZ regime, leading to distinct scaling properties of the fronts before and after  formation of these large scale sawtooth shapes. Once the final static state is reached, the front exhibits a larger roughness than the one determined during the transient propagation, or excluding these inclined static regions.

\subsection{A.\;\;\;\;Additional experimental  scaling analysis}

In the pinned phase with negative $F$, the front exhibits a higher roughness exponent once the final static state is reached. Fig.~\ref{WL} shows the corresponding $w(\Delta x)$ functions: left, determined from the sawtooth shape fronts; center, during the transient propagation, excluding the inclined static regions. The temporal height  fluctuations determined during the transient regime for the same values of $F$ are displayed on Fig.~\ref{WT} left end center respectively.
\begin{figure}
\includegraphics[width= 15.5cm]{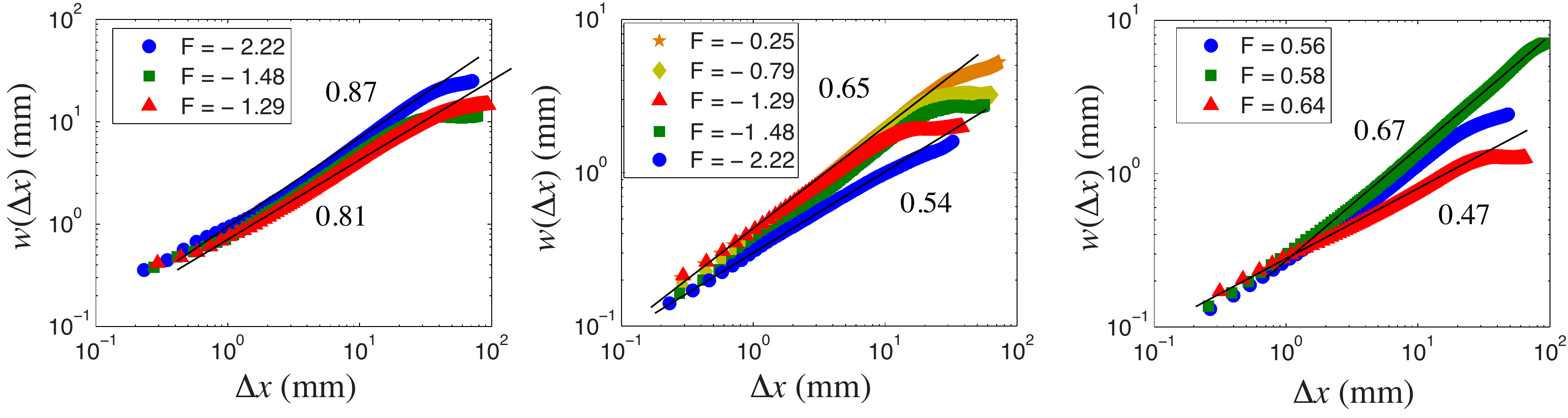}
\caption{Left and center: roughness of the fronts in the negative qKPZ regime. Left: determined for the final sawtooth static fronts, center: determined during the transient propagation and excluding the frozen portions of the front. Right: roughness of the fronts in the positive qKPZ regime for upstream propagating fronts.}
\label{WL}
\end{figure}
\begin{figure}
\includegraphics[width= 15.5 cm]{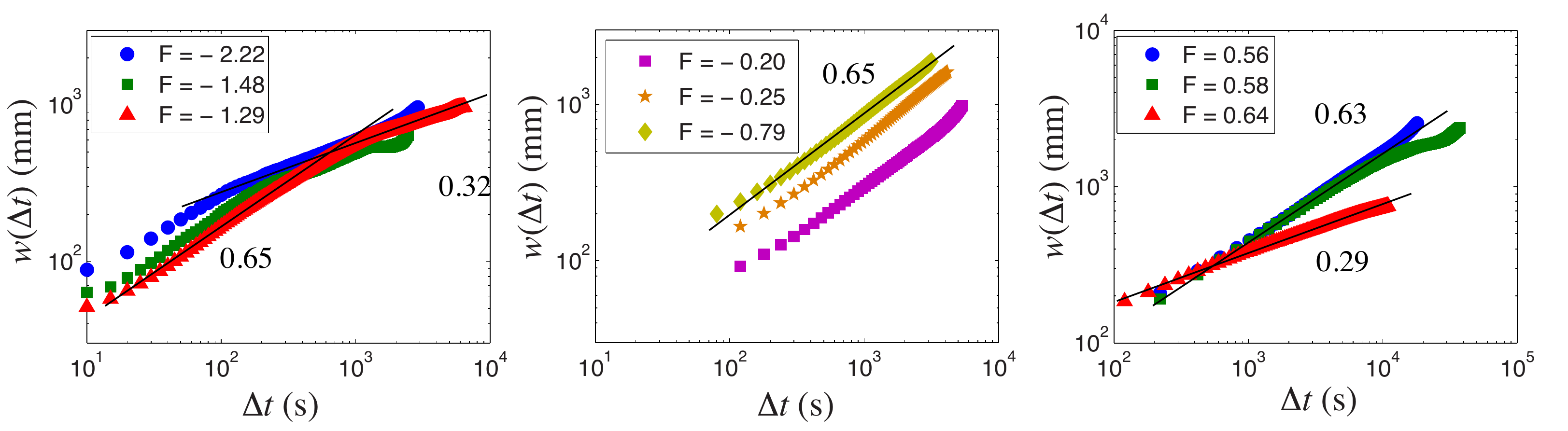}
\caption{Temporal fluctuations in the qKPZ regime determined during the transient propagation of the front. Left: close to to $\Fcd$ excluding the sawtooth shape frozen portions, center: for intermediate values of negative $F$, and right: close to to $\Fcu$.}
\label{WT}
\end{figure}
When $F$ becomes less negative, i.e.\ $F \gtrsim -1$, both the sawtooth size and inclination become smaller, and no more distinction is necessary between the scalings of the static state and the transiently moving state.

For the positive qKPZ regime, one can note on Figs.~\ref{WL} (right) and \ref{WT} (right) that the thermal KPZ exponents are recovered for $F=0.64$.
For higher value of $F$, when $F\rightarrow 1+f_0=1.38$, the mean flow velocity vanishes and the noise is too weak to generate height fluctuations. This area corresponds to the hatched area in Fig. 4 in the manuscript. As a consequence, the temporal fluctuations become uncorrelated: $\beta \rightarrow 0$, and the roughness of the front stays constant at $\alpha\simeq 0.8$.\\

\subsection{B.\;\;\;\;Additional numerical  scaling analysis}

\begin{figure}
\includegraphics[width= 10 cm]{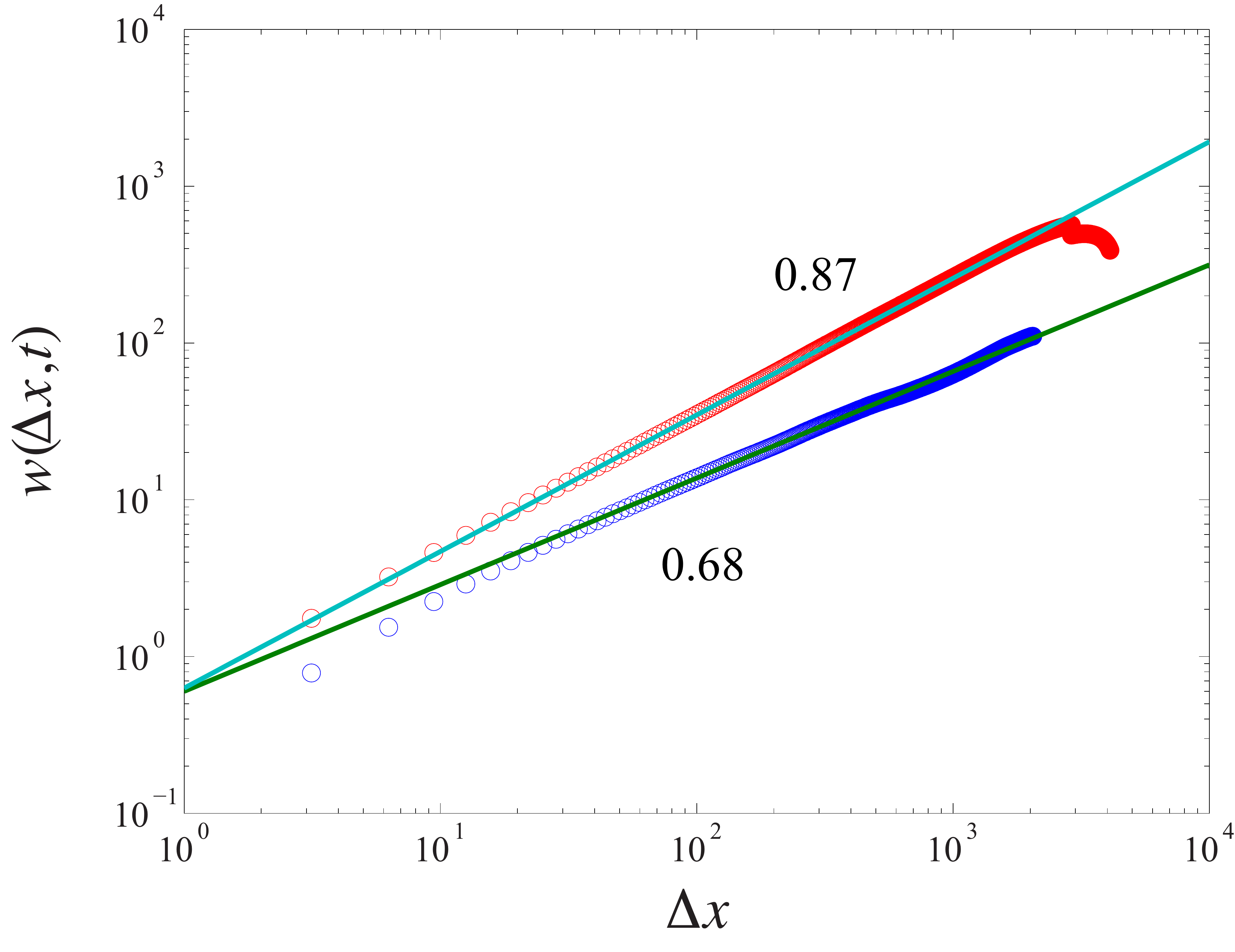}
\caption{Numerical front roughness of the final static fronts at $\Flau \simeq \Flcp$, blue, and $\Flau \simeq \Flcm$, red.}
\label{WLsimu2}
\end{figure}
As in the experiments, measuring a roughness exponent of the static front is  more complex, and the same distinction between  static fronts and  moving fronts is necessary. As indicated on Fig.~\ref{WLsimu2}, when $\Flau \rightarrow \Flcm$, the roughness of the static front $\alpha =0.87 \pm 0.05$ becomes close to one indicating the characteristic sawtooth shape. 
Fig.~\ref{WLsimu2} also shows the roughness of the front close to $\Flcp$, with the value $\alpha =0.68 \pm 0.05$ in agreement with the positive qKPZ equation. In this regime, it is interesting to note that there is a discrepancy between the roughness measured in the moving regime ($\alpha=0.75\pm0.05$) and the one measured in the static regime, which should be compared to  Directed Percolation \cite{amaral95}.
Far from the two transitions when $\Flau < \Flcm$ or $\Flau > \Flcp$, the two exponents are  close to the ``thermal'' KPZ ones. Indeed  at $\Flau=-6$, $\alpha=0.55 \pm 0.05$ and $\beta = 0.32 \pm 0.05$ and at $\Flau=1$, one has $\alpha=0.45 \pm 0.05$ and $\beta = 0.32 \pm 0.05$.

\section{V.\;\;\;\;Region of low $F$}

As $\Flau$ is decreased further on Fig. 5 in the manuscript, the roughness of the numerical fronts drops to zero at $\Flau \simeq -0.5$ and then increases again as $F$ approaches the second transition at $\Flcm$. We can note here that, deep inside the pinned phase, we expect the final front shape to depend on the initial condition. Both in the experiments and the numerics, when $\Flau \rightarrow 0$, the front moves only by a small amount before getting pinned. In fact, it is worth noting that the Poisson point disorder models (first layer PNG as in \cite{szabo01} or in a more realistic form in \cite{gueudre14}) predict that the final pinned configuration has the same roughness as the initial one. Hence in the numerics it is expected that the roughness drops to zero as the initial condition is flat, whereas in the experiments the initial fronts are never  flat as they have already traveled through the glass beads before the flow is switched on.
When we get closer to $\Flcm$, the front moves a longer distance before getting pinned, and thus one recovers a more universal behavior as $\Flau \rightarrow \Flcm$.

\section{VI.\;\;\;\;Control parameter definition}

As a control parameter, we define $F = \frac{\bar U + V_\chi}{V_\chi} + f_0$ in the experiments and $\Flau = \frac{\bar U + V_\chi}{V_\chi} + \hat{f}_0$ in the simulations. One can note that the value $F=1+f_0$ is quite singular, since then the noise vanishes but not the driving force.
In the porous medium, it is quite unclear if one should use $V_\chi$ or $V_0=0.8\Vch \pm 0.5 \mu$m/s (\emph{i.e.} including or not the effect of tortuosity/porosity) to determine $F$. Here, we are using $\Vch$ to define $F$ as suggested by the eikonal approximation. Indeed, the parameter $\Vch$ was shown to lead to the right estimation of the sawtooths angle when the fronts are static \cite{atis13}. In the simulations, porosity and tortuosity are both equal to one, and  $V_0=V_\chi$.
\begin{figure}[t]
\includegraphics[width= 7 cm]{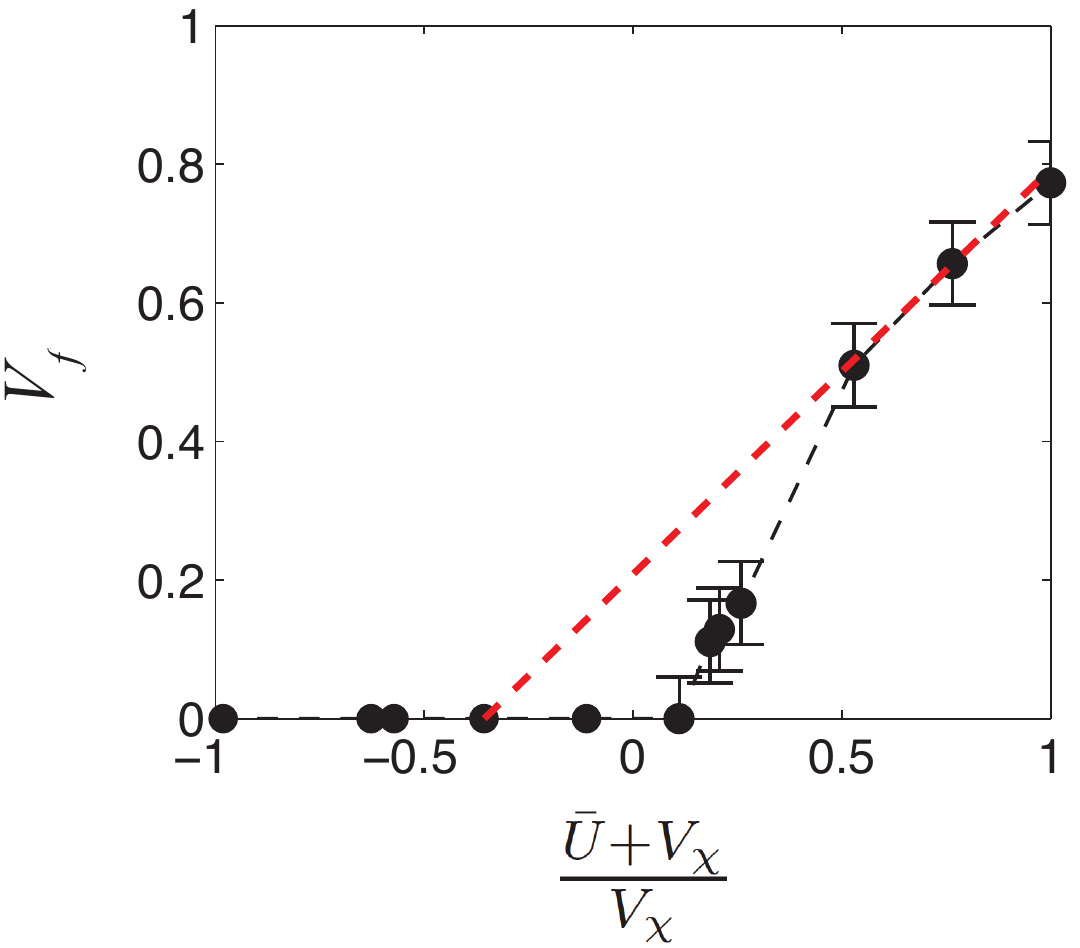}
\includegraphics[width= 7 cm]{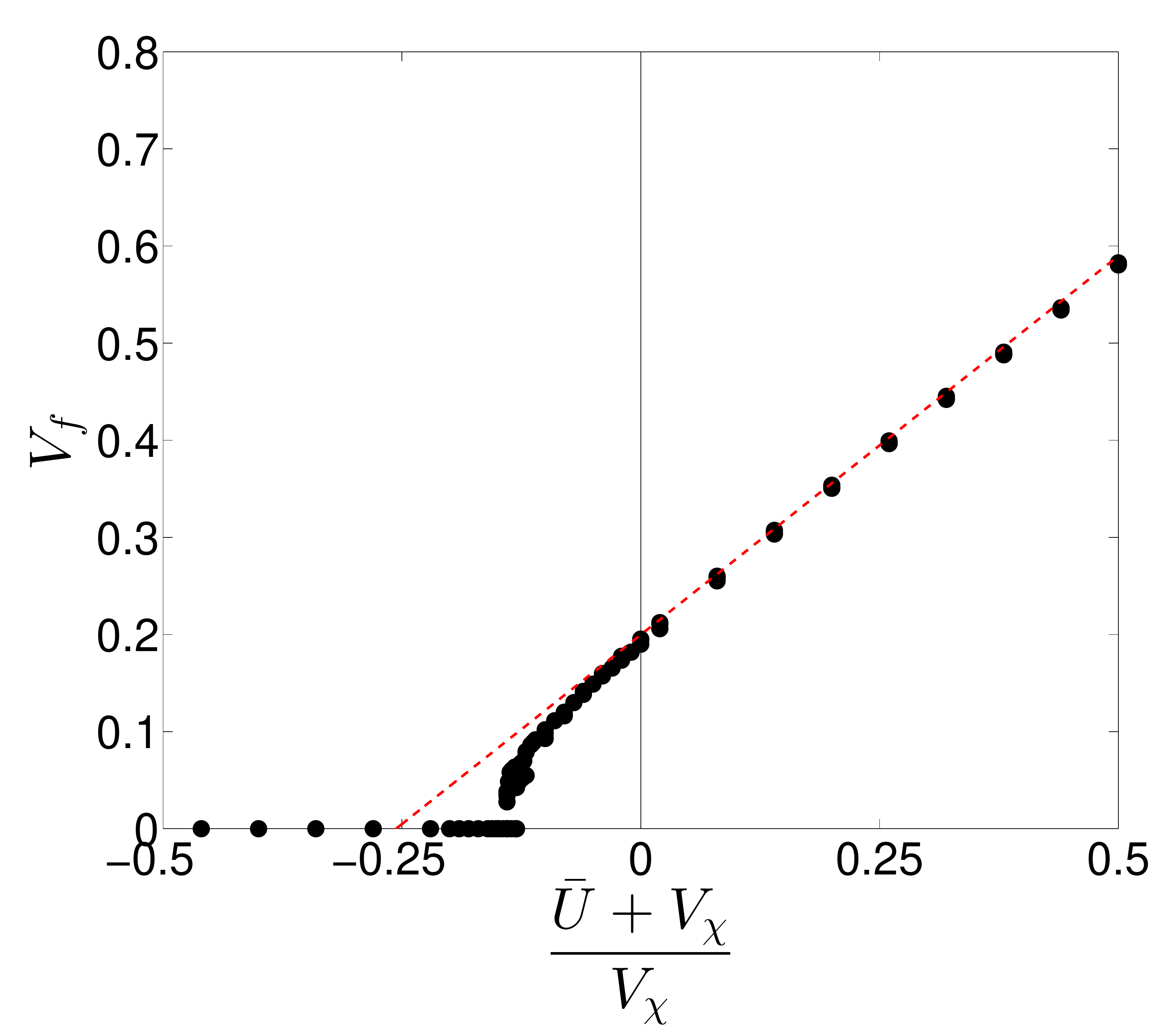}

\caption{Front mean velocity versus the relative mean flow rate: $(\Um + \Vch )/ \Vch$. Dashed red lines indicate the linear fit with the right moving branch and its intersection with the $x$ axis determines the value of $f_0$. Left: $f_0 = 0.38$ in experiments, and  right: $f_0=0.256$ in simulations.}
\label{VF_zoom}
\end{figure}

By analogy with the qKPZ behavior, the constant $f_0$ is determined by fitting the right velocity branch far from its pinning threshold (see Figs.~\ref{VF_zoom}). As described below, we attribute this constant value to the average roughness of the front.
A closer inspection of the Fig.~\ref{VF_zoom} right, shows that it can be fitted with a power-law $\Vf(\hat F) \sim ( \hat F - \Flcp )^{0.8 \pm 0.05} $, with  $\Flcp \simeq 0.095 \pm 0.015$  as shown on Fig.~\ref{Vf_FFc}. This exponent is slightly larger than the theoretically predicted value $0.66$ \cite{amaral95}. One can note from Fig.~\ref{VF_zoom} left, that this exponent is hard to determine experimentally as the number of available data points close to $\Fcu$ is too  small. The reason for this limitation lies in  experimental uncertainties, constraining the number of distinct $F$ to approximately 4 in this region.

\begin{figure}[t]
\includegraphics[width= 7 cm]{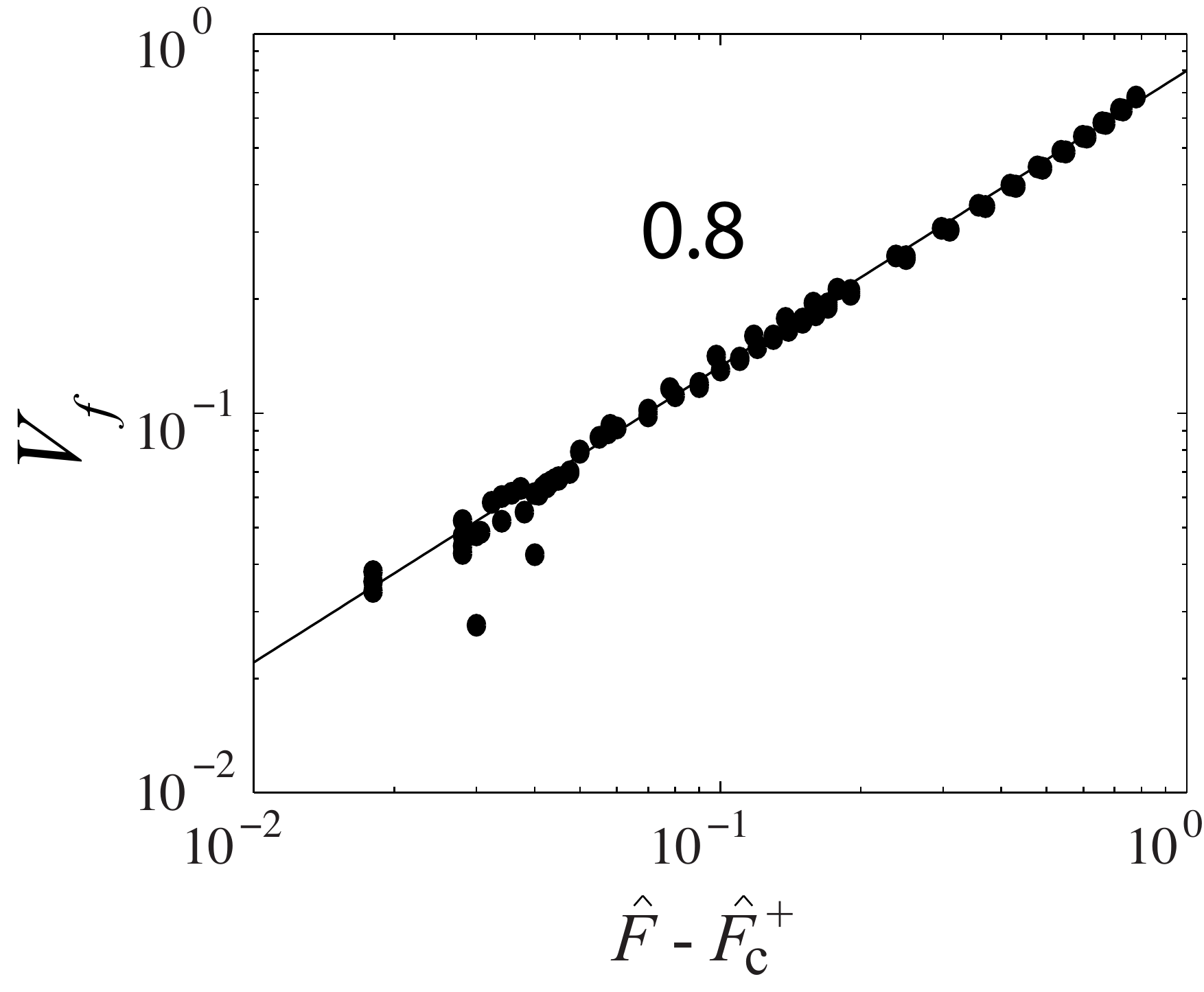}
\caption{Numerical front velocity versus $\Flau - \Flcp$.}
\label{Vf_FFc}
\end{figure}

In the numerics, below the first transition, $\Flau<\Flcm$, and beyond the second one, $\Flau>\Flcp$, the velocity $\Vf(\Flau)$  is almost linear with  $\Vf = 0.9 \Flau + 0.52 $ and $\Vf = 0.8 \Flau $, respectively.
Both branches have different slopes and differ from the prediction of the KPZ equation ($v/f \sim 1$ at  large $f$). Similarly, in the experiments, we note that both for negative $F<\Fcd$ and for positive $\Fcu<F<1.38$, the  linear branches have a different slope than one: $\Vf=0.55F - 0.15$ and $\Vf=0.56F$ respectively. In addition, for the supportive flow regime in the experiments, this slope becomes $\Vf = 1.12 F-0.85$.
We attribute this behavior to a nonlinear interaction between the chemical reaction and the flow field, which enhances mixing and front velocity as studied in \cite{xin00,edwards02,edwards06}. Indeed, since flow heterogeneities necessarily enhance the mixing and thus the reaction rate, one can expect at first order in $\epsilon=\bar U/v_\chi$ to have an effective chemical velocity $V_\chi \rightarrow  V_\chi ( 1 +  a |\epsilon|)$ leading to an average front velocity $\Vf = (1-a) \bar U + V_\chi$ for adverse flow ($\epsilon<0$), and $\Vf = (1+a) \bar U + V_\chi$ for supportive flow ($\epsilon>0$).

\section{VII.\;\;\;\;From the eikonal approximation to the qKPZ equation}

For a thin front, the local front velocity follows the eikonal equation:
\begin{equation}
\vec \Vf \cdot \vec n =  \Vch + D_m \kappa + \vec U(\vec r) \cdot \vec n \;\text{,}
\end{equation}
where $\vec n$ is the normal of the front, $\kappa$ the curvature and $\vec U(\vec r)$ the local flow velocity. Indeed, this equation is similar to the ``flux-line model'' of Kardar \cite{kardar98}
where the chemical velocity plays the role of the Lorentz force and the 
disordered flow of the random force. After projection the eikonal approximation yields
\begin{eqnarray*}
&& \frac{\partial h}{\partial t} = \sqrt{1 +s^2} \left[ D_m \partial_x^2 h / (1  + s^2) ^{3/2} +  \Vch + \left(\Um + \delta U_y(\vec r) - s \delta U_x(\vec r) \right)/\sqrt{1 + s^2}  \right] \;\text{,} \\
&&{\rm with} \;\; s=\nabla h \;\;{ \rm and } \;\; \vec U(\vec r) = \bar U \vec e_y + \delta  \vec U(\vec r)\;\text{.}
\end{eqnarray*}
Considering that the flow is highly anisotropic, $\delta U_x \ll \delta U_y$, this gives
\begin{equation}
\frac{\partial h}{\partial t} \simeq  \frac{D_m \nabla^2 h} {1  + (\nabla h)^2} + V_\chi\sqrt{1  + (\nabla h)^2} +\Um + \delta U_y(\vec r).
\end{equation}
Finally, assuming small gradients and neglecting higher-order terms, this leads to the qKPZ equation,
\begin{equation}
\frac{\partial h}{\partial t} \simeq  D_m \nabla^2 h + \frac{\Vch}{2} (\nabla h)^2  + \delta U_y(\vec r) + \Um + \Vch\;\text{.}
\end{equation}
After normalizing by $V_\chi$, we can write the qKPZ equation with the following parameters:
\begin{eqnarray}
\nu & = & l_\chi = D_m / \Vch\\
\lambda & = & 1 \\
\bar \eta &\equiv& \delta U_y\\
f & = & \frac{\Um + \Vch }{\Vch}\ .
\end{eqnarray}
Near the transition at $\Fcd$, the small-gradient limit may not be quantitatively accurate in this region as the sawtooth-pattern slope becomes large.
In this regime, since pinning occurs where the flow field is slower than the chemical reaction, the transition results from the competition of two effects: (i) the probability to find a region of low flow velocity decreases as $F\rightarrow -\infty$, (ii) when the ``sawtooths" get sharper, the probability to find new pinning points increases.
Based on extreme-value statistics and the PNG model \cite{szabo01}, a more precise scenario has been  proposed in \cite{gueudre14}.

\section{VIII.\;\;\;\;Disorder measurements}

In the experiment $ \bar D\sim \Vf$ and $\Vf$ is proportional to $F$ for large mean flow velocities, i.e.\ $|F|> 2.5$. The expected KPZ noise $D \simeq \bar D/\Vf \sim \bar D/ F$ becomes then almost independent of $F$. 
$\bar D$ can be deduced experimentally from the local flow-velocity fluctuations. The width $\sigma(F)\sim \bar D$ of the corresponding flow-velocity PDF, at a given value of $F$, can be estimated with tracer transport experiments \cite{atis13}. Fig.~\ref{sigma} a) shows the value of $\sigma(F)$ determined from tracer front dispersion experiments; it shows that $\sigma$ is proportional to $F$ for both signs of $F$ and $|F|>2$. 
The dependence of $\sigma/F$ with $F$ is shown on Fig.~\ref{sigma} b). One can note that $\sigma /F \sim \bar{D}/F \simeq \rm const$ for  $F>2$, with a different value depending on the the sign of $F$.
\begin{figure}[t]
\includegraphics[width= 9cm]{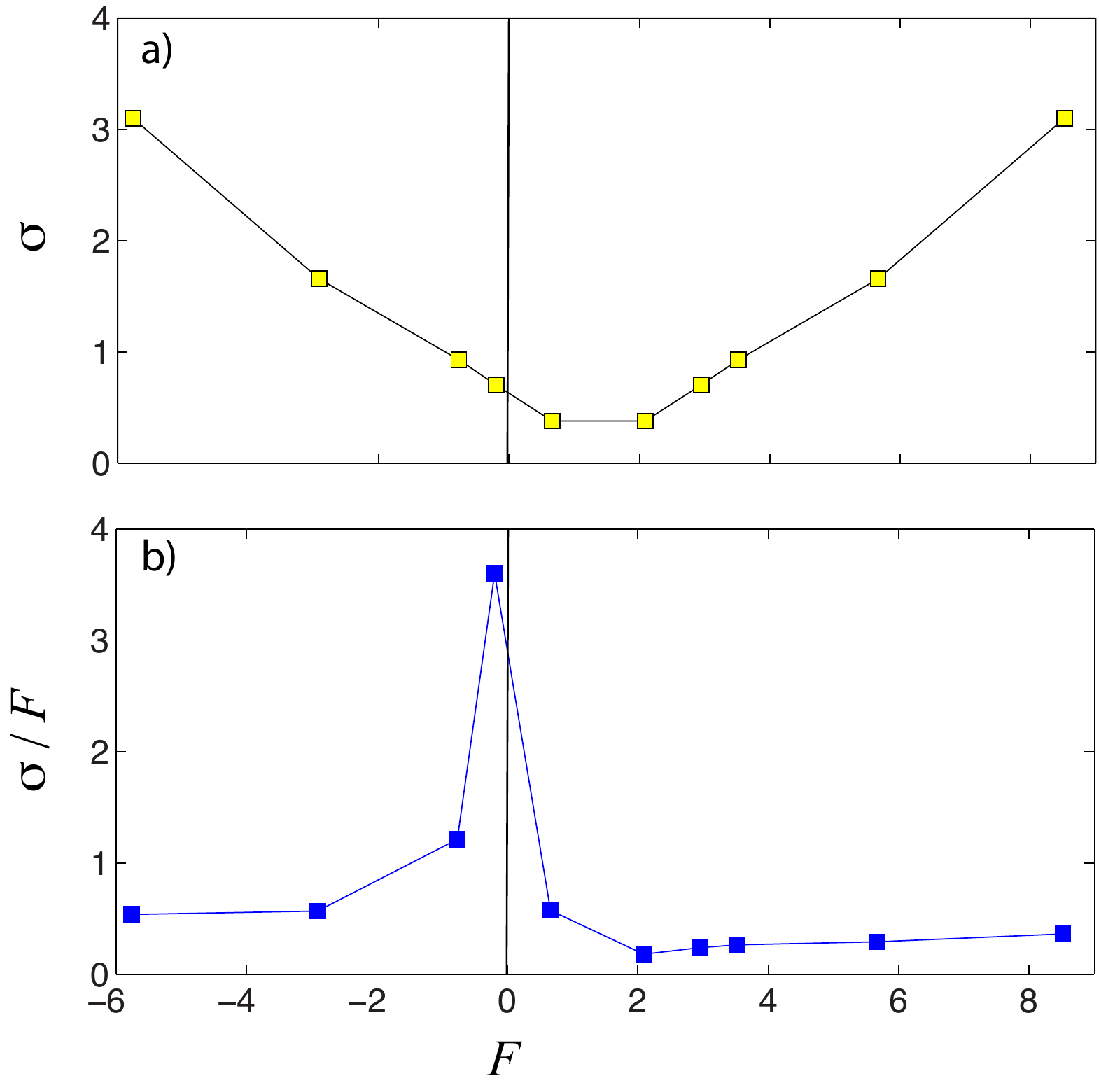}
\caption{a) The dependence on $F$ of the width $\sigma$ of the flow-velocity  PDF determined from tracer transport experiments. b) the dependence of the estimated KPZ noise $D$  on $F$: $\sigma/F \sim \bar D /\F \simeq D$.}
\label{sigma}
\end{figure}

\newpage

\section{IX.\;\;\;\;Supplemental Movies}

\subsection*{Supplemental movie 1}

Reaction front propagation in the absence of a flow, i.e.\ at $F=1+f_0\simeq1.38$.\\
\textbf{\url{http://www.fast.u-psud.fr/~atis/Movie1_Q0.avi}}\\

\subsection*{Supplemental movie 2}

Upward propagating reaction fronts at large flow rate, for positive mean flow orientation at $F\simeq 3.16$. In this configuration both the chemical wave velocity and the mean flow velocity are oriented towards the top.\\
\textbf{\url{http://www.fast.u-psud.fr/~atis/Movie2_Q025_up.avi}}\\

\subsection*{Supplemental movie 3}

Downward propagating reaction fronts at large flow rate, for negative mean flow orientation at $F\simeq - 4.34$. In this configuration the chemical wave velocity is oriented toward the top and the mean flow velocity toward the bottom.\\
\textbf{\url{http://www.fast.u-psud.fr/~atis/Movie3_Q08_down.avi}}\\

\subsection*{Supplemental movie 4}

Upward propagating reaction front at low flow rate, for negative mean flow orientation at $F\simeq 0.58$. In this configuration the chemical wave velocity is oriented towards the top and the mean flow velocity towards the bottom.\\
\textbf{\url{http://www.fast.u-psud.fr/~atis/Movie4_Q011_down.avi}}\\

\subsection*{Supplemental movie 5}

Zero velocity reaction fronts for negative mean-flow orientation at $F\simeq 0.2$. The chemical wave velocity is oriented towards the top and the mean-flow velocity towards the bottom. The front propagates at the chemical reaction velocity and stops when the flow is turned on.\\
\textbf{\url{http://www.fast.u-psud.fr/~atis/Movie5_Q015_down.avi}}\\

\subsection*{Supplemental movie 6}

Downward propagating reaction fronts at low flow rate, for negative mean flow orientation at $F\simeq - 1.25 $. In this configuration the chemical-wave velocity is oriented towards the top and the mean-flow velocity towards the bottom.
\textbf{\url{http://www.fast.u-psud.fr/~atis/Movie5_20_down_SSB.avi}}\\


\end{widetext}


\begin{thebibliography}{34}
\bibitem{footnote1}
We define: $w(\Delta x) = \left<\sqrt{\left< [h(x,t)-\left<h\right>_{\Delta x}]^2\right>_{\Delta x}}\right>_{x,t}$; here $\left<\right>_{\Delta x}$ represents an average over a length $\Delta x$, and  $\left<\right>_{x,t}$ a translational average over $x$, as well as a temporal average over time $t$, in the steady state. Temporal fluctuations are quantified by $w(\Delta t) = \left<\sqrt{\left< [h(x,t)-\left<h(t)\right>_x]^2\right>_{\Delta t}}\right>_{x,t}$, where $\Delta t$ is a time window. Both quantities exhibit power laws, $ w(\Delta x) \sim \Delta x^\alpha$ for $\Delta x\ll\ell^*(t)$, and $w(\Delta t)  \sim  \Delta t^\beta$ for $\Delta t\ll\ t^*(L)$, with $\ell^*(t)$ and $t^*(L)$ the spatial and temporal correlation lengths.
\end{thebibliography}
\end{document}